\documentclass[10pt,twocolumn,english,prl,showpacs,nofootinbib]{revtex4}
\usepackage{graphicx}
\usepackage{amsmath}
\usepackage{amssymb}
\usepackage{amsfonts}
\usepackage{color}
\usepackage{bm}



\newcommand{\nn}{\nonumber}
\newcommand{\newc}{\newcommand}
\newc{\be}{\begin{equation}}
\newc{\ee}{\end{equation}}
\newc{\ba}{\begin{eqnarray}}
\newc{\ea}{\end{eqnarray}}
\newc{\bea}{\begin{eqnarray*}}
\newc{\eea}{\end{eqnarray*}}

\def \om {\Omega_{\rm m}}

\def \Geff {G_{\rm eff}}

\def \fs8 {$f\sigma_8$ }
\def \OO {$\mathcal{O}$}

\begin{document}
\title{A novel null test for the $\Lambda$CDM model with growth-rate data}

\author{Savvas Nesseris}
\email{savvas.nesseris@uam.es}

\affiliation{Departamento de F\'isica Te\'orica and Instituto de F\'isica Te\'orica, \\
Universidad Aut\'onoma de Madrid IFT-UAM/CSIC,\\ $28049$ Cantoblanco, Madrid, Spain}

\author{Domenico Sapone}
\email{domenico.sapone@uam.es}

\affiliation{Departamento de F\'isica Te\'orica and Instituto de F\'isica Te\'orica, \\
Universidad Aut\'onoma de Madrid IFT-UAM/CSIC,\\ $28049$ Cantoblanco, Madrid, Spain}
\affiliation{Departamento de F\'isica, Universidad de Chile,\\
Blanco Encalada 2008, Santiago, Chile}

\pacs{95.36.+x, 98.80.-k, 98.80.Es}

\begin{abstract}

Current and upcoming surveys will measure the cosmological parameters with an extremely high accuracy.
The primary goal of these observations is to eliminate some of the currently viable cosmological models created to explain the late time accelerated expansion (either real or only inferred). However, most of the statistical tests used in cosmology have a strong requirement: the use of a model to fit the data. Recently there has been an increased interest on finding tests that are model independent, i.e. to have a function that depends entirely on observed quantities and not on the model, see for instance \cite{clarksonbassett}. In this letter we present an alternative consistency check at the perturbative level for a homogeneous and isotropic Universe filled with a dark energy component. This test makes use of the growth of matter perturbations data and it is able to not only test the homogeneous and isotropic Universe but also, within the framework of a Friedmann-Lema\^itre-Robertson-Walker Universe, if the dark energy component is able to cluster, if there is a tension in the data or if we are dealing with a modification of gravity.

\end{abstract}
\maketitle

\section{Introduction}

The observed accelerated expansion of the late-time Universe has led cosmologists to revise the concept of gravity either by introducing a new form of matter in the Universe called dark energy \cite{sapone-review} or by modifying directly the laws of gravity \cite{tsujikawa}. Within the framework of Friedmann-Lema\^itre-Robertson-Walker (FLRW) cosmology, we can account for a phase of accelerated expansion by simply introducing a cosmological constant ($\Lambda$), even though this model gives rise to severe coincidence and fine-tuning problems, observations still predict that the Universe is filled by a dark energy component that has the same characteristics of the cosmological constant \cite{planck}, \cite{sdss2014}.

However, these observations say little about either the possible time evolution of dark energy, which might lead dark energy to cluster, or about modified gravity models. Even though the two classes of models can be arbitrarily alike \cite{ks}, it is still important to be able to discard some of the models we now have.

In light of the future experiments, planned to collect an enormous amount of data, it would be interesting to find a test that is able to confirm our assumptions. Consistency checks are usually model independent tests able to determine if any of the underlying assumptions made are violated. Their big advantage is that they tend to be quite generic and that they have to be true at all redshifts.

In this letter we present a new test for the $\Lambda$CDM model starting from the differential equation that governs the growth of matter density contrast in a homogenous and isotropic Universe in the limit of small scale approximation and zero dark energy perturbations\footnote{The cosmological constant has zero perturbations.}. From this differential equation we construct a null test that is able to verify (or falsify) the assumptions used to obtain the evolution of matter density contrast, by using direct observations. The idea is similar to the test presented in \cite{clarksonbassett} (and tested with recent data in \cite{curvature}), where the authors considered a test for background quantities (the Hubble parameter and the comoving distance) to verify the assumption of homogeneity in the Universe; another consistency check is the $Om$ statistic of Shafieloo et al. \cite{Sahni:2008xx}. This work looks at this direction and here we give a new null test that is able to verify whether the Universe is homogeneous and isotropic under the assumption of Einstein gravity and if matter is the only component able to form bound objects. If the result is different from what would be expected, even at a single redshift, then one of the core assumptions made is incorrect.

For completeness, we also test our consistency check with recent data, such as measurements of the Hubble parameter and the growth data. Anticipating the result, we find that the data confirm at $1\sigma$ that the Universe is FLRW filled with a cosmological constant.

\section{The null test for the growth-rate}

The growth of matter in the Universe under the assumption of homogeneity and isotropy with zero dark energy perturbations, is governed by the second order differential equation:
\be
\delta''(a)+\left(\frac3a + \frac{H'(a)}{H(a)}\right)\delta'(a) -\frac{3\Omega_{m_0}\delta(a)}{2 a^5 H(a)^2/H_0^2}=0,
\label{ode}
\ee
where $H(a)$ is the Hubble parameter, $\Omega_{m_0}$ is the matter density contrast today and $H_0$ is the Hubble constant. Equation (\ref{ode}) is valid only under the assumption of FLRW Universe with a cosmological constant (which has no perturbations). Any deviations from this scenario will imply a modification to the Eq.~(\ref{ode}).
For instance, if we assume that the dark energy model is not a cosmological constant then the dark energy is able to cluster and these perturbations will influence the evolution of the matter density contrast, see \cite{saponeDEP, saponeDEP2, Sapone:2012nh, Sapone:2013wda}. If we drop altogether the assumption of a dark energy component and we assume that the theory of gravity is modified, then Eq.~(\ref{ode}) itself has to be modified and the evolution of the matter density contrast will be different, see \cite{tsujikawa1}.

Furthermore, if the assumption of homogeneity and isotropy is broken, then Eq.~(\ref{ode}) will also be modified, see \cite{Alonso:2012ds, Alonso:2010zv}.

\subsection{Lagrangian formalism}

To find a null test that involves the evolution of the matter density, we will make use of the Lagrangian formalism. The first step is to find a Lagrangian for Eq.~(\ref{ode}) and, with the help of the Noether's theorem,
to find an associated conserved quantity. If we assume that the Lagrangian can be written as $\mathcal{L}=\mathcal{L}(a,\delta(a),\delta'(a))$, where $a$, $\delta(a)$ and $\delta'(a)$ are the ``time", the  generalized position and velocity variables of the system respectively, then the Euler-Lagrange equations are:
\be
\frac{\partial \mathcal{L}}{\partial \delta}-\frac{d}{da}\frac{\partial \mathcal{L}}{\partial \delta'}=0\,. 
\label{ELeqs}
\ee
So, let us assume a Lagrangian of the form
\ba
\mathcal{L}&=&T-V \nn \\
T&=&\frac12 f_1(a,H(a))\delta'(a)^2 \nn \\
V&=&\frac12 f_2(a,H(a))\delta(a)^2 \nn
\ea
where the $f_1$ and $f_2$ are arbitrary functions that need to be determined so that the resulting equation after implementing the Euler-Lagrange Eq.~(\ref{ELeqs}) is exactly Eq.~(\ref{ode}). Therefore, substituting the former Lagrangian in the Euler-Lagrange Eq.~(\ref{ELeqs}) and comparing the result with Eq.~(\ref{ode}) we are able to get the two functions $f_1$ and $f_2$ and consequently to build the Lagrangian $\mathcal{L}$ of the system:
\be
\mathcal{L}=\frac12 a^3 H(a)/H_0\delta'(a)^2+\frac{3\om}{4a^2 H(a)/H_0}\delta(a)^2\,.
\label{Lag}\\
\ee
It is easy to see that substituting Eq.~(\ref{Lag}) into Eq.~(\ref{ELeqs}) results exactly to Eq.~(\ref{ode}).

Now that we have a Lagrangian we can use Noether's theorem to find a conserved quantity that will be later translated to the null test. So, if we have an infinitesimal transformation $\textbf{X}$ with a generator
\ba
\textbf{X}&=&\alpha(\delta) \frac{\partial }{\partial \delta}+ \frac{d\alpha(\delta)}{da}\frac{\partial }{\partial \delta'} \\
\frac{d\alpha(\delta)}{da}&\equiv&\frac{\partial \alpha}{\partial\delta}\delta'(a)=\alpha'(a),
\ea
such that for the Lie derivative of the Lagrangian we have $L_X \mathcal{L}=0$, then
\be
\Sigma=\alpha(a)\frac{\partial \mathcal{L}}{\partial \delta'}\label{const1}
\ee
is a constant of ``motion" for the Lagrangian of Eq.~(\ref{Lag}), see Ref.~\cite{Capozziello:2007iu} for an application of Noether's theorem in Scalar-Tensor cosmology and more details. From Eq.~(\ref{const1}) we get that
\be
\Sigma= a^3 H(a)/H_0 \alpha(\delta)\delta'(a),
\ee
while from the Lie derivative we also obtain:
\be
\alpha(a)=c\;e^{-\int_{a_0}^a\frac{3\om  \delta(x)}{2x^5 H(x)^2/H_0^2\delta'(x)}dx},
\ee
where $c$ is an integration constant and $a_0$ can be chosen for convenience to be either 0 or 1. Then the constant becomes
\be
\Sigma= a^3 H(a)/H_0 \delta'(a)\;e^{-\int_{1}^a\frac{3\om \delta(x)}{2x^5 H(x)^2/H_0^2\delta'(x)}dx},
\label{const2}
\ee
where we have redefined $\Sigma$ to absorb $c$. The constant $\Sigma$ in general is $\Sigma\equiv a_0^3 H(a_0)/H_0\delta'(a_0)= \delta'(1)$, and Eq.~(\ref{const2}) can be normalized to unity. We have checked numerically the validity of Eq.(\ref{const2}) for several different cosmologies and values of the parameters. Also, it is easy to confirm that taking the derivative of Eq.~(\ref{const2}) with respect to the scale factor $a$, results in Eq.~(\ref{ode}) as expected.

However, surveys cannot provide observations of $\delta(a)$ directly, but instead they measure $f\sigma_8(a)\equiv f(a) \sigma_8(a)$, where $\sigma_8(a)=\sigma_8(a=1)\frac{\delta(a)}{\delta(a=1)}$, hence we need to properly transform Eq.~(\ref{const2}) to be able to test it directly with observations. Taking into account $f(a)\equiv \frac{d ln \delta}{dlna}$, we have that
\be
f\sigma_8(a)\equiv f(a) \sigma_8(a)=\xi a \delta'(a),
\ee
where $\xi\equiv \frac{\sigma_8(a=1)}{\delta(a=1)}$, which gives
\be
\delta(a)=\delta(1)+\int_{1}^a\frac{f \sigma_8(x)}{\xi x}dx,
\ee
where we chose the initial condition to be at $a_0=1$ without any loss of generality.
Then, Eq.~(\ref{const2}) can be written as
\be
\mathcal{O}(z)=a^2 \frac{H(a)}{H_0}\frac{f\sigma_8(a)}{f\sigma_8(1)}e^{-\frac{3}{2}\Omega_{m_0}H_0^2\int_{1}^a\frac{\sigma_{8_0}+\int_{1}^x\frac{f\sigma_8(y)}{y}dy}{x^4H(x)^2f\sigma_8(x)}dx},
\label{nullf}
\ee
where we rewrite $\sigma_{8_0}=\sigma_8(a=1)$ for simplicity. It is clear that Eq.~(\ref{nullf}) have to be constant for all redshifts $z$ and moreover $\mathcal{O}(z)$ has to be equal to $1$ and
any deviation from unity might highlight any of the four different scenarios:
\begin{itemize}
  \item A deviation from the FLRW Universe.
  \item Non zero dark energy perturbations
  \item A deviation from GR, ie modified gravity, such as $f(R)$ models.
  \item Tension between the $H(z)$ and $f\sigma_8$ data.
\end{itemize}

We would also like to add that the consistency check given by Eq.~(\ref{nullf}) can be farther generalized by introducing an effective Newtonian constant $\Geff$ that accounts for modified gravity models or for perturbations in the dark energy sector. However, this goes beyond the goal of this letter as we are more concerned with finding a test for the $\Lambda$CDM model, which is still the reference cosmological model for observations. Therefore, we leave the generalization of the null test of Eq.~(\ref{nullf}) to an upcoming publication,  \cite{nulltest-2}.

\section{Data analysis}
In this Section we will use Eq.~(\ref{nullf}) to construct a new null test primarily for the growth data. Since this equation only holds for GR with the FLRW metric, deviations point to either new physics or tension in the data. Clearly, the LHS of Eq.~(\ref{nullf}) has to be constant for all redshifts $z$ regardless of the background DE model. Therefore, any deviation implies that there is either a deviation from GR and the FLRW model or a systematic in the data.

We reconstruct the null test $\mathcal{O}(z)$ by using different cosmological measurements. In order to reconstruct Eq.~(\ref{nullf}) we need four independent observables: the Hubble parameter $H(z)$, the $f\sigma_8(z)$, $\sigma_8(z=0)$ and $\Omega_{m_0}h^2$. In the current analysis we use the $f\sigma_8$ data from different experiments (collected by Refs \cite{Basilakos:2013nfa} and \cite{taddei}) and the Hubble parameters values found in Moresco et al. \cite{moresco_etal} (see also \cite{curvature}). Instead of  $\sigma_8(z=0)$ and $\Omega_{m_0}h^2$, we use the values from \cite{tojeiro} and \cite{planck}, respectively.

\begin{figure}
\centering
\vspace{0cm}\rotatebox{0}{\vspace{0cm}\hspace{0cm}\resizebox{0.48\textwidth}{!}{\includegraphics{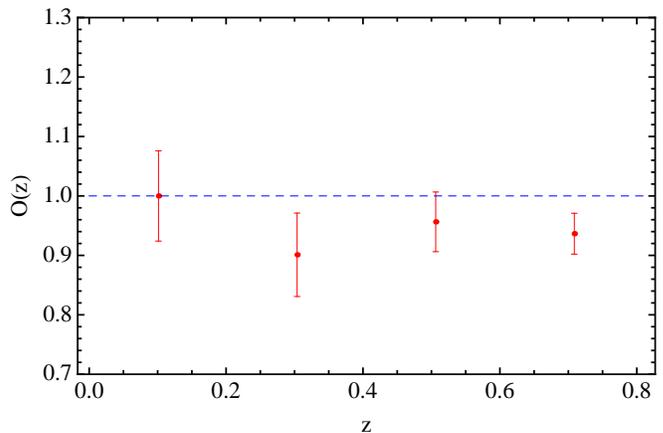}}}
\caption{The results for the null test with the binning method for the actual data, using $H(z)$ and $f\sigma_8(z)$.
\label{fig:binrealvsmock-null}}
\end{figure}

The binning technique to measure \OO$(z)$ consists of evaluating it in several redshift bins by directly computing the $H(z)$ values measured from passively evolving galaxies data and by using the $f\sigma_{8}(z)$ values measured by different experiments. The Hubble parameter catalog contains $n_{H}=19$ data spanned from redshift $0.1$ up to $z=1.75$, whereas the growth rate catalog contains $n_{f\sigma_8}=17$ data points from $z=0.02$ up to $z=0.8$.
Since the growth measurements reach only up to  $z=0.8$ we are forced to discard the last 8 data points for $H(z)$ (as we want to avoid having too wide bins). Because the number of data for both catalogs is quite small the choice of the bins is quite restricted. We decided to opt for $4$ bins equally spaced such that there are at least two data points for each bin and we evaluate the observables at the mean redshift of the bins. It is important to notice that, in order to maintain consistency for the null test $\mathcal{O}(z)$, we need to evaluate quantities at the same redshift. We show the results in Fig.~\ref{fig:binrealvsmock-null}. As can be seen from the figure, the actual data, even if poor at the moment, are in agreement already at $1\sigma$ with a FLRW Universe, filled by a cosmological constant. The errors of the null test $\mathcal{O}(z)$ are in average of the order of $\sim 0.06$.

It is interesting to note that the data manifest a value of $\mathcal{O}(z)$ less than unity indicating perhaps that matter does not cluster as much as expected. To understand this we can have a closer look at Eq.~(\ref{nullf}). Its dependence on the growth rate is $\sim f\sigma_8(z)\exp{\left[-1/\left( f\sigma_8(z)\right)\right]}$, so if matter clusters more or, also, more rapidly, i.e. if $f\sigma_8(z)$ increases than the quantity $\mathcal{O}(z)$ decreases because any variation in the exponent will affect strongly the final results. We would also like to point out that this is a test for the growth itself; in fact, any change of the Hubble parameter will affect equally the growth and the $\mathcal{O}=1$ at all redshifts. This is quite obvious, but it is interesting to clarify it again as the earlier discussion about the dependence of $\mathcal{O}(z)$ on $f\sigma_8$ focuses on the latter only. For the sake of completeness, we should also stress that the data we used to reconstruct the null test are still few and as a result, the deviation from unity found is not statistically significant.

With upcoming surveys, like Euclid and DES, it will be possible to substantially reduce the errors on the observed quantities like the Hubble parameter and the growth rate, but more importantly we will have a larger number of data points available.


\section{Conclusions}\label{sec:conclusions}

In this {\em letter} we have presented a new test of $\Lambda$CDM cosmology that makes use of the growth-rate data. As shown, this test is able to verify (or falsify) not only the assumptions of homogeneity and isotropy of the Universe but also if we are dealing with modified gravity models or with a dark energy component which can cluster. What does  $\mathcal{O}(z)\neq 1$ really tell us? A deviation from unity of the null test $\mathcal{O}(z)$ even at one single redshift will imply that Eq.~(\ref{ode}) is not the correct equation of motion for the matter density contrast. Hence, statistically significant deviations from unity mean out that we live in a Universe which is not homogeneous and isotropic \cite{Alonso:2012ds}, \cite{Alonso:2010zv}; or that we need to go beyond the cosmological constant scenario and start to account for perturbations also in the dark energy sector \cite{sapone-review}; or that some sophisticated modified gravity model should be considered \cite{tsujikawa}.

It is fair to mention that this null test alone is not capable of pointing out the correct direction to follow as any deviation from unity might be caused by a different effect; however, together with other null tests, like the $\Omega_{K}(z)$ presented in \cite{clarksonbassett} which is able to test the assumptions of homogeneity and isotropy of the Universe, our growth null test will be extremely useful.

\section*{Acknowledgments}

It is a pleasure to thank Juan Garc\'ia-Bellido for interesting discussions. SN and DS acknowledge financial support from the Madrid Regional Government (CAM) under the program HEPHACOS S2009/ESP-1473-02, from MICINN under grant AYA2009-13936-C06-06 and Consolider-Ingenio 2010 PAU (CSD2007-00060), as well as from the European Union Marie Curie Initial Training Network UNILHC Granto No. PITN-GA-2009-237920. We also acknowledge the support of the Spanish MINECO's ``Centro de Excelencia Severo Ochoa" Programme under Grant No. SEV-2012-0249.

{}

\end{document}